\documentclass[journal,twocolumn]{IEEEtran}

\usepackage{amsmath,amsfonts,amssymb,amsthm}
\usepackage{cite}
\usepackage{graphicx}
\usepackage{xcolor}
\usepackage[small]{caption}
\usepackage{epstopdf}		
\usepackage{algorithm}
\usepackage{algpseudocode}

\newcommand{\myT}{{\mathsf{T}}} 

\title{Robust Single- and Multi-Pinching Antenna  Systems Under User Location Uncertainty}
\author{Hao Feng, Ebrahim Bedeer, Ming Zeng, Xingwang Li, Wanming Hao and Dingzhu Wen
    \thanks{H. Feng is with Hunan Institute of Engineering, Xiangtan, China, and Donghua University, Shanghai, China as well as Laval University, Quebec city, Canada (email: 1219001@mail.dhu.edu.cn).}

    \thanks{E. Bedeer is with the Department of Electrical and Computer Engineering, University of Saskatchewan, Saskatoon, SK, Canada (email: e.bedeer@usask.ca).}
    
    \thanks{M. Zeng is with the Department of Electrical and Computer Engineering,  Laval University, Quebec City, Canada (email: ming.zeng@gel.ulaval.ca).}

    \thanks{X. Li is with School of Physics and Electronic Information Engineering, Henan Polytechnic University, Jiaozuo, China (email: lixingwang@hpu.edu.cn).}

    \thanks{W. Hao is with the School of Electrical and Information Engineering, Zhengzhou University, Zhengzhou 450001, China (e-mail: iewmhao@zzu.edu.cn).}

    \thanks{D. Wen is with the School of Information Science and Technology, ShanghaiTech University, Shanghai 201210, China (e-mail: wendzh@shanghaitech.edu.cn).}

}

\begin{document}
	\maketitle

	\begin{abstract}
		Pinching antenna (PA) systems have recently emerged as a promising architecture for reconfigurable wireless communications by enabling flexible antenna placement along a dielectric waveguide. However, existing works typically assume perfect knowledge of user locations, which is impractical in real systems where location estimation errors are inevitable. In this paper, we investigate robust power allocation and antenna placement for PA systems under user location uncertainty. We consider both single-antenna and multi-antenna configurations, where the true user locations are unknown but lie within bounded uncertainty regions. For the single-antenna case, we adopt a worst-case robust design and leverage the S-procedure to transform the joint power allocation and antenna placement problem into a convex semidefinite program (SDP), ensuring that quality-of-service (QoS) constraints are satisfied for all possible user locations. For the multi-antenna case, we address the additional challenges arising from the superposition of channel components from multiple antennas by developing an efficient numerical procedure to evaluate the worst-case channel gain. Then, we derive a closed-form solution for optimal power allocation and develop a block coordinate descent algorithm to optimize antenna placement. Simulation results show that the proposed framework provides robustness to location uncertainty while achieving power consumption close to that of outage-based benchmark schemes.
	\end{abstract}
	
	
\section{Introduction}\label{sec:Introduction}
The rapid proliferation of Internet of Things (IoT) applications, including smart cities, industrial automation, and intelligent transportation systems, has led to an unprecedented growth in the number of connected devices \cite{Al-Fuqaha_2015, Guo_2021}. These systems require wireless communication solutions that are energy-efficient, scalable, and capable of adapting to dynamic environments. In this context, pinching antenna (PA) systems have recently emerged as a promising architecture for flexible and reconfigurable wireless communications \cite{Atsushi_22, yang2025, liu2026survey,  zeng2025_WCM, wijewardhana2025}. By ``pinching'' radiating elements onto a dielectric waveguide, PA systems allow electromagnetic signals to propagate within the waveguide and radiate at controllable spatial locations. This architecture introduces a new degree of freedom, i.e., antenna position optimization, enabling dynamic spatial reconfiguration without requiring complex radio-frequency hardware. Compared to conventional antenna arrays, PA systems offer reduced hardware complexity, improved energy efficiency, and enhanced adaptability, making them attractive for next-generation wireless networks \cite{ Zhao_TWC26, Xiao_COMML25,  Ouyang_COMML25,  Zhao_TCOM25, fu2025, zeng2025EE}. 

Most existing works on PA systems assume that user locations are perfectly known at the transmitter \cite{ding2024, Zeng_COMML25, Tegos_2025}. Under this assumption, the wireless channel can be deterministically modeled, and accurate channel state information (CSI) can be obtained to support resource allocation and system optimization. However, in practical systems, user locations are typically obtained through positioning or sensing techniques, which inevitably introduce estimation errors due to noise, limited resolution, and environmental dynamics. Since the PA channel is inherently dependent on the geometric relationship between antennas and users, even small localization errors can lead to significant mismatch in channel modeling. This mismatch may result in degraded performance or violation of quality-of-service (QoS) requirements, highlighting the need for robust PA system design under user location uncertainty.

Our prior works \cite{zeng2025robust, feng2026} have taken initial steps toward addressing this issue. In \cite{zeng2025robust}, user location uncertainty is modeled as a bounded region, and a geometric approach is employed to determine the transmit power required to satisfy outage probability constraints. In \cite{feng2026}, a probabilistic framework is adopted by modeling user locations as Gaussian random variables, enabling a refined characterization of uncertainty and corresponding power allocation strategies. While these works provide useful insights, they are limited to single-antenna (single-PA) systems and rely on heuristic optimization methods, such as particle swarm optimization (PSO), for antenna placement.

In addition, robust resource allocation under imperfect CSI has been extensively studied in multi-input multi-output (MIMO) and reconfigurable intelligent surface (RIS)-aided communication systems \cite{Sun_TCOM18, Hao_TCOM24}. In such systems, uncertainty is typically addressed via worst-case optimization. However, these techniques cannot be directly applied to PA systems due to the unique structure of the channel, which is explicitly governed by the geometric relationship between antenna positions and user locations. In particular, the coupling between antenna placement and channel propagation introduces additional complexity that is absent in those systems.

Furthermore, while probabilistic approaches (e.g., outage-constrained optimization) can reduce conservatism, they often require accurate knowledge of the distribution of location errors, which may not be available in practice. In contrast, worst-case robust designs provide deterministic guarantees and are particularly appealing for mission-critical IoT applications requiring strict QoS assurance.

Motivated by these limitations, this paper develops a unified and tractable framework for robust power allocation and antenna placement in PA systems under user location uncertainty, covering both single-antenna and multi-antenna cases. The main contributions of this work are summarized as follows:
\begin{itemize}
    \item {\bf{Robust single-antenna design via convex reformulation:}} For the single-PA case, we adopt a worst-case robustness approach and employ the S-procedure to guarantee that QoS constraints are satisfied for all possible user locations within a prescribed uncertainty region. By jointly optimizing transmit power and antenna position, we reformulate the problem as a convex semidefinite program (SDP), thereby eliminating the need for heuristic search methods such as PSO and enabling efficient computation of globally optimal solutions. Simulation results show that the proposed worst-case design achieves performance comparable to probabilistic outage-based approaches, providing strong robustness guarantees without sacrificing efficiency.
    \item {\bf{Multi-antenna extension with worst-case channel characterization:}} For the general multi-PA case, we address the additional challenges arising from the superposition of channel components from multiple antennas. We first propose an efficient numerical method to compute the worst-case channel gain over the uncertainty region. This enables us to derive a closed-form solution for optimal power allocation, given antenna positions. The antenna placement problem is then solved using a block coordinate descent (BCD) algorithm with multiple initializations to obtain a locally optimal solution with manageable computational complexity.
\end{itemize}


The paper is organized as follows: Section II introduces the system model and formulates the robust optimization problem, whereas Section III presents the proposed solutions for both the single- and multi-antenna cases. Section IV presents the simulation results while Section V concludes the paper. 
	
\begin{figure}
\centerline{\includegraphics[width=1\linewidth]{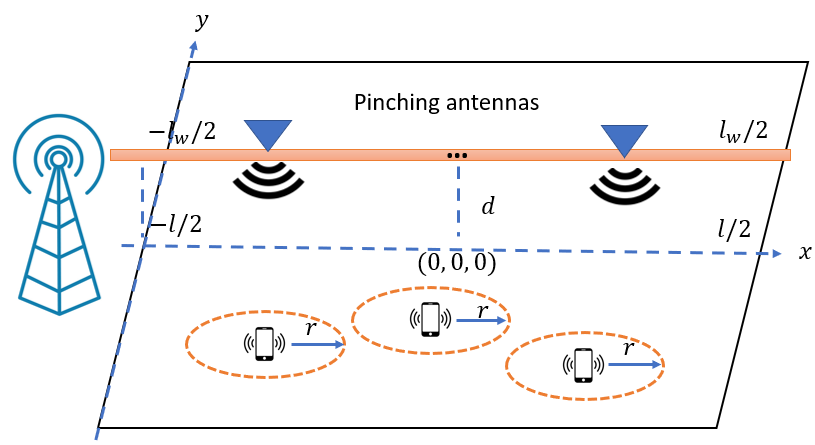}}
\caption{System model for the considered pinching antenna system with bounded uncertainty regions.} \label{fig1}
\end{figure}
	
\section{System Model}\label{sec:system_model}
We consider a downlink communication scenario in which an access point (AP) serves $K$ single-antenna users through $N$ PAs deployed on a single dielectric waveguide, as shown in Fig. \ref{fig1}. The $K$ users are uniformly distributed  on the ground plane within a rectangular area of dimension $\ell_{\rm{x}} \times \ell_{\rm{y}}$ meters, and we assume that the center of the rectangular area is located at the origin of the Cartesian coordinate system. The dielectric waveguide, of length $\ell_{\rm{w}}$, is located along the $x$-axis at height $d$ above the ground plane; hence, the location of the $n$th PA, $n = 1, ..., N,$ is denoted as $\tilde{\mathbf{v}}_n = [v_n, 0, d]^\myT$ where $v_n \in [-\ell_{\rm{w}}/2, \ell_{\rm{w}}/2]$. We assume the feed point is located at $\tilde{\mathbf{v}}_0 = [-\ell_{\rm{w}}/2, 0, d]^\myT$. The true location of the $k$th user, $k = 1, ..., K,$ is $\mathbf{u}_k = [u_{x,k}, u_{y,k}, 0]^\myT$; however, in this paper, we assume that this true location is not known exactly due to inaccuracies in the positioning of the users, and only an estimate  $\hat{\mathbf{u}}_k = [\hat{u}_{x,k}, \hat{u}_{y,k}, 0]^\myT$ is available such that the location estimation error, $\Delta \mathbf{u}_k = \mathbf{u}_k - \hat{\mathbf{u}}_k$, is bounded by a circle of radius $r_k$, i.e., 
	\begin{IEEEeqnarray}{RCL}
		\Delta \mathbf{u}_k^\myT \Delta \mathbf{u}_k &\leq& r_k^2.
	\end{IEEEeqnarray}
	
The combined channel $h_k$ from the AP to the user $k$ can be expressed as
	\begin{IEEEeqnarray}{RCL}
		h_k(\mathbf{u}_k, \mathbf{v})
		& = &
		\sum_{n = 1}^N
		\frac{\sqrt{\eta}}{{d}_{k,n}}
		\exp\left(
		- j 2 \pi
		\left(
		\frac{{d}_{k,n}}{\lambda}
		+ \frac{v_n + \ell_{\rm{w}}/2}{\lambda_g}
		\right)
		\right), \nonumber \\ &&
		\label{eq:ch} \IEEEeqnarraynumspace
	\end{IEEEeqnarray}
	where ${d}_{k,n} = \lVert \mathbf{u}_k - \tilde{\mathbf{v}}_n \rVert$ is the distance from the $n$th PA to the $k$th user, $v_n + \ell_{\rm{w}}/2 = \lVert \tilde{\mathbf{v}}_0 - \tilde{\mathbf{v}}_n \rVert$ is the distance from the feed point to the $n$th PA, and $\mathbf{v} = [v_1, ..., v_N]^\myT$ which is the vector that contains the $x$-axis coordinates of all PAs, $\eta = \lambda^2/(16 \pi^2)$ is the free space path loss coefficient, $\lambda$ is the wavelength, and $\lambda_g = \lambda/n_{\rm{eff}}$ where $n_{\rm{eff}}$ is the waveguide effective refractive index. As can be seen in \eqref{eq:ch}, the combined channel $h_k$ is a function of the optimization variable $\mathbf{v}$ that includes the $x$-coordinates of all the PAs and the unknown true user $k$ location, i.e., $\mathbf{u}_k$. 
    
The channel model in eq. \eqref{eq:ch} captures two key propagation mechanisms. The first term represents the guided-wave propagation inside the dielectric waveguide, where the phase shift depends on the effective wavelength $\lambda_g$. The second term corresponds to free-space radiation from each PA element to the user, characterized by the distance-dependent attenuation and phase rotation. 

It is worth noting that, unlike conventional antenna arrays where signals are directly radiated from the feed network, PA systems introduce an additional propagation stage inside the waveguide. This results in a hybrid propagation model that combines guided and radiative components, thereby enabling more flexible spatial signal shaping through antenna position optimization.

We assume the AP serves the $K$ users using time division multiple access (TDMA) with equal time slots allocated to each user; hence, there is no inter-user interference in the downlink transmission. We also assume the PAs operate in the equal power mode, i.e., the power assigned  to a given user is equally distributed among the $N$ antennas. That being said, the received signal at the $k$th user is written as
	\begin{IEEEeqnarray}{RCL}
		y_k &=& \sqrt{\frac{p_k}{N}} 	h_k(\mathbf{u}_k, \mathbf{v}) s_k + w_k, \label{eq:sig_rec}
	\end{IEEEeqnarray}	
	where $p_k$ is the power allocated by the AP to user $k$, $s_k$ is the unit-power information symbol of user $k$, and $w_k$ is the additive white Gaussian noise (AWGN) with zero mean and variance $\sigma_k^2$. Hence, the achievable rate of user $k$ is given as
	\begin{IEEEeqnarray}{RCL}
		R_k = \frac{1}{K} \log_2 \left(1 + \frac{p_k |	h_k(\mathbf{u}_k, \mathbf{v})|^2}{N \sigma_k^2}\right). \label{eq:rate}
	\end{IEEEeqnarray}
	Please note that the terms $1/N$ and $1/K$ in \eqref{eq:sig_rec} and \eqref{eq:rate} reflect the equal power emitted by each PA and the equal time slot durations of each user, respectively.
	
	As mentioned earlier, positioning of users may not be fully accurate, and hence, only an estimate of each user location is available at the AP, which may degrade the user rate and possibly lead to service outage. In this paper, we allow a certain outage to occur at user $k$, which is formally written as
	\begin{IEEEeqnarray}{RCL}
		{\rm{Pr}}(R_k < R_{k,\min}) & \leq & \epsilon_k, \label{eq:rate_const}
	\end{IEEEeqnarray}
	where $R_{k,\min}$ and $\epsilon_k$ represent the minimum user rate and the maximum allowable outage at user $k$, respectively. That being said, the optimization problem to minimize the total transmit power by the AP is formulated as
    \begin{subequations}
    \begin{IEEEeqnarray}{RCL}
		\mathbb{P}1: \quad	\min_{p_k, \mathbf{v}} &\quad & \sum_{k=1}^K p_k  \label{eq:op1_ob}\\ 
		{\rm{s.t.}} & \quad & {\rm{Pr}}(R_k < R_{k,\min})  \leq  \epsilon_k,  \quad \forall k,\label{eq:op1_a}\\ 
		& \quad & v_n - v_{n-1} \geq \lambda/2, \quad \forall n, \label{eq:op1_b}\\
		& \quad & -\ell_{\rm{w}}/2 \leq v_n \leq \ell_{\rm{w}}/2, \quad \forall n, \label{eq:op1_c}\\
		& \quad & p_k \geq 0, \label{eq:op1_d}
	\end{IEEEeqnarray}   
    \end{subequations}
	where the constraint \eqref{eq:op1_b} prevents the coupling effect between PAs, and the constraint \eqref{eq:op1_c} ensures that the PAs are located on the waveguide.
	
\section{Proposed Solution}\label{sec:sol}
We first treat the single-PA case, i.e., $N=1$, and provide a worst-case solution using the S-procedure. Following this, we treat the multiple-PA case, i.e., $N>1$. 
	
\subsection{Single-PA Case: Worst Case Solution}\label{sec:sproc_distance}
For $N=1$, the squared channel magnitude gain reduces to the free-space form
	\begin{IEEEeqnarray}{rcl}
	|h_k(\mathbf{u}_k,v)|^2 \;=\; \frac{\eta}{\lVert \mathbf{u}_k - \tilde{\mathbf{v}}\rVert^2},
	\end{IEEEeqnarray}
	where $\tilde{\mathbf v}=[v,0,d]^\top$ denotes the single PA coordinate. Using \eqref{eq:rate}, the per-user signal-to-noise ratio (SNR) requirement equivalent to $R_k\ge R_{k,\min}$ is
	\begin{equation}\label{eq:snr_ming2}
		\frac{\eta p_k}{\lVert\mathbf{u}_k-\tilde{\mathbf{v}}\rVert^2}\ge \gamma_{k,\min},
	\end{equation}
	where $\gamma_{k,\min}\triangleq \sigma_k^2\big(2^{K R_{k,\min}}-1\big)$. The probabilistic constraint \eqref{eq:rate_const}  becomes
	\begin{equation}\label{eq:prob_snr}
		\Pr\!\Big( \eta p_k / \lVert\mathbf{u}_k-\tilde{\mathbf{v}}\rVert^2  < \gamma_{k,\min}\Big)\le\epsilon_k.
	\end{equation}
	A conservative, yet deterministic, approach to handle the probabilistic constraint \eqref{eq:prob_snr} enforces \eqref{eq:snr_ming2} to be satisfied for all possible user locations in the uncertainty ball $\{\mathbf{u}_k:\Delta\mathbf{u}_k^\top\Delta\mathbf{u}_k\le r_k^2\}$. This robust requirement can be written as
	\begin{IEEEeqnarray}{rCl}
		q_1(\mathbf{u}_k) &\triangleq& \mathbf{u}_k^\top\mathbf{u}_k - 2\hat{\mathbf{u}}_k^\top\mathbf{u}_k + \hat{\mathbf{u}}_k^\top\hat{\mathbf{u}}_k - r_k^2 \le 0, \label{eq:q1}\\
		q_2(\mathbf{u}_k) &\triangleq& \eta p_k - \gamma_{k,\min}\lVert\mathbf{u}_k-\tilde{\mathbf{v}}\rVert^2 \ge 0. \label{eq:q2}
	\end{IEEEeqnarray}
	By the S-procedure \cite{boyd1994inequalities }, since $q_1(\hat{\mathbf u}_k)=-r_k^2<0$, i.e., $q_1(\hat{\mathbf u}_k)$ is strictly negative at $\mathbf{u} = \hat{\mathbf{u}}$; hence, Slater condition holds \cite{boyd2004convex}. This implies  there exists $\lambda_k\ge0$ such that
	\begin{IEEEeqnarray}{rcl}
	q_2(\mathbf{u}_k)+\lambda_k q_1(\mathbf{u}_k)\ge0,\quad\forall\mathbf{u}_k.
	\end{IEEEeqnarray}
	Collecting quadratic, linear, and constant terms in \(\mathbf u_k\) yields the quadratic form
	\begin{IEEEeqnarray}{RcL}
	( \lambda_k - \gamma_{k,\min} ) \mathbf{u}_k^\top\mathbf{u}_k
	+ 2( \gamma_{k,\min}\tilde{\mathbf{v}} - \lambda_k\hat{\mathbf u}_k)^\top\mathbf{u}_k \hspace{1cm}
	\nonumber \\ \hspace{1cm} + \eta p_k - \gamma_{k,\min}\tilde{\mathbf{v}}^\top\tilde{\mathbf{v}} + \lambda_k(\hat{\mathbf u}_k^\top\hat{\mathbf u}_k - r_k^2)
	\ge 0. \IEEEeqnarraynumspace
	\end{IEEEeqnarray}
	This quadratic inequality is nonnegative for all $\mathbf u_k$ if and only if the corresponding symmetric block matrix is positive semidefinite. Since the location uncertainity is on the ground plane, define the 2D ground-plane user vectors $\bar{\mathbf u}_k=[u_{x,k},u_{y,k}]^\top$, $\hat{\bar{\mathbf u}}_k=[\hat u_{x,k},\hat u_{y,k}]^\top$, and $\bar{\mathbf v}=[v,0]^\top$. Then the linear matrix inequality (LMI) condition can be written compactly as \eqref{eq:M_ming_correct} at the top of the next page.
	\begin{figure*}[!t]
		\begin{equation}\label{eq:M_ming_correct}
			\mathbf{M}_k\big(p_k,\lambda_k,v\big)
			=
			\begin{bmatrix}
				(\lambda_k - \gamma_{k,\min})\mathbf{I}_2
				&
				\gamma_{k,\min}\,\bar{\mathbf v} - \lambda_k\,\hat{\bar{\mathbf u}}_k
				\\[0.6ex]
				\big(\gamma_{k,\min}\,\bar{\mathbf v} - \lambda_k\,\hat{\bar{\mathbf u}}_k\big)^{\!\top}
				&
				\eta p_k - \gamma_{k,\min}\big(\bar{\mathbf v}^{\top}\bar{\mathbf v} + d^2\big)
				+ \lambda_k\big(\hat{\bar{\mathbf u}}_k^{\top}\hat{\bar{\mathbf u}}_k - r_k^2\big)
			\end{bmatrix}
			\succeq \mathbf{0},
		\end{equation}
		\hrulefill
	\end{figure*}
 Note that the top-left block of  \eqref{eq:M_ming_correct} being $(\lambda_k-\gamma_{k,\min})\mathbf I_2$ implies $\lambda_k\ge\gamma_{k,\min}$ whenever the LMI holds; therefore one may enforce $\lambda_k\ge0$ and include the stronger bound $\lambda_k\ge\gamma_{k,\min}$ explicitly for numerical stability.
	
	The dependence of $\mathbf M_k\big(p_k,\lambda_k,v\big)$ on $v$ is affine except for the scalar quadratic $\bar{\mathbf v}^\top\bar{\mathbf v}=v^2$ appearing in the bottom-right entry. We introduce a slack variable $t_k$ to upper bound that quadratic term as
	\begin{equation}\label{eq:slack_t}
		t_k \ge \bar{\mathbf v}^\top\bar{\mathbf v} = v^2,
	\end{equation}
	to obtain an affine LMI $\bar{\mathbf M}_k(p_k,\lambda_k,v,t_k)\succeq0$ as shown in \eqref{eq:M_with_t_correct} on the top of the next page.
	\begin{figure*}[!t]
		\begin{equation}\label{eq:M_with_t_correct}
			\bar{\mathbf{M}}_k\big(p_k,\lambda_k,v,t_k\big)
			=
			\begin{bmatrix}
				(\lambda_k - \gamma_{k,\min})\mathbf{I}_2
				&
				\gamma_{k,\min}\,\bar{\mathbf v} - \lambda_k\,\hat{\bar{\mathbf u}}_k
				\\[0.6ex]
				\big(\gamma_{k,\min}\,\bar{\mathbf v} - \lambda_k\,\hat{\bar{\mathbf u}}_k\big)^{\!\top}
				&
				\eta p_k - \gamma_{k,\min}\big(t_k + d^2\big)
				+ \lambda_k\big(\hat{\bar{\mathbf u}}_k^{\top}\hat{\bar{\mathbf u}}_k - r_k^2\big)
			\end{bmatrix}
			\succeq \mathbf{0}.
		\end{equation}
		\hrulefill
	\end{figure*}
	Finally, using the LMI \eqref{eq:M_with_t_correct} and the epigraph constraint \eqref{eq:slack_t} we obtain the following convex SDP for the single-PA worst-case robust optimization problem
    \begin{subequations}
	\begin{IEEEeqnarray}{RCL}
		\mathbb{P}2: \quad		\min_{ p_k, v, \lambda_k, t_k } &\quad& \sum_{k=1}^K p_k \\
		{\rm s.t.} && \bar{\mathbf M}_k\big(p_k,\lambda_k,v,t_k\big)\succeq\mathbf 0,\quad \forall k, \label{P2_lmi}\\
		&& t_k \ge \bar{\mathbf v}^\top\bar{\mathbf v}, \quad \forall k, \label{P2_t}\\
		&& \lambda_k \ge \gamma_{k,\min},\ \ \forall k, \quad\label{P2_lambda_nonneg}\\
		&& -\ell_{\rm w}/2 \le v \le \ell_{\rm w}/2,\\
		&& p_k\ge 0,\quad \forall k.
	\end{IEEEeqnarray} \label{P2_final}
    \end{subequations}
 The problem $\mathbb P2$ is an SDP and can be solved with standard solvers \cite{grant2014cvx}.

{\bf{Remark:}}
The SDP formulation in eq. \eqref{P2_final} reveals that the worst-case robust design effectively enlarges the feasible region of the uncertainty set by enforcing QoS constraints for all possible realizations. Compared to probabilistic formulations, this approach may appear conservative; however, as shown in the simulation results, the associated performance loss is negligible. This indicates that the geometric structure of the PA channel inherently limits the worst-case degradation, making robust design particularly effective in this context.

	\subsection{Multiple-PA Case}\label{sec:N_greater_1}

We now consider the general case with multiple PAs ($N>1$).  We first compute
the worst-case channel gain over the uncertainty region as follows. 
For a fixed PA placement $\mathbf v$, we define the {worst-case squared channel magnitude}
for user $k$ as
\begin{equation}\label{eq:worstcase_def_repeat}
	m_k(\mathbf v)\;=\;\min_{\|\Delta\mathbf{u}\|\le r_k}
	\bigl|h_k(\hat{\mathbf u}_k+\Delta\mathbf u,\mathbf v)\bigr|^2.
\end{equation}

The optimization problem in eq. \eqref{eq:worstcase_def_repeat} is non-convex due to the nonlinear dependence of the channel on the user location. Moreover, the objective function may exhibit multiple local minima within the uncertainty region due to the superposition of signals from multiple antennas. This makes it challenging to obtain a closed-form expression for $m_k(\mathbf v)$, necessitating efficient numerical approximation methods.

It is worth noting that, due to the radial symmetry of the uncertainty region, the worst-case channel gain often occurs on or near the boundary of the uncertainty set. This observation motivates the boundary sampling strategy adopted in the following, which significantly reduces the search space while maintaining high accuracy. The proposed boundary sampling followed by a local refinement approach is given as follows:

\begin{itemize}
	\item We sample the circle $\|\Delta\mathbf u\|=r_k$
	uniformly at $L$ angles $\theta_\ell=2\pi(\ell-1)/L$, $\ell=1,\dots,L$,
	and evaluate
	\begin{IEEEeqnarray}{rcl}
		f_\ell \;=\; \bigl|h_k(\hat{\mathbf u}_k + r_k[\cos\theta_\ell,\sin\theta_\ell]^\top,\mathbf v)\bigr|^2.
	\end{IEEEeqnarray}
	Retain the smallest value of $f_\ell$ and its angle as a candidate.
	
	\item We use the obtained best boundary candidate together
	with several interior starting points as initial points for a local constrained
	minimizer (e.g., MATLAB's \texttt{fmincon}) with  the disk
	constraint
	$\Delta\mathbf u^\top\Delta\mathbf u\le r_k^2$. Then, we keep
	the smallest objective value of \eqref{eq:worstcase_def_repeat}, and denote the numerical approximation as $\hat m_k(\mathbf v)$.
\end{itemize}
To avoid numerical
instability when $\hat m_k(\mathbf v)$ is extremely small, we introduce a
positive floor $\varepsilon$ and define the
{safe} worst-case gain
\begin{equation}\label{eq:floor_repeat}
	\tilde m_k(\mathbf v) \;=\; \max\{\hat m_k(\mathbf v),\,\varepsilon\}.
\end{equation}

The robust SNR requirement (i.e., enforcing the rate constraint for every
realization inside the uncertainty ball) is equivalent to
\begin{equation}\label{eq:power_safe_repeat}
	p_k \;\ge\; \frac{N \gamma_{k,\min}}{\tilde m_k(\mathbf v)}.
\end{equation}
That being said, for a fixed PA placement $\mathbf v$, the optimal power allocation
minimizing total transmit power is the linear program
\begin{IEEEeqnarray}{rcl}
\min_{p_k\ge0}\;\sum_{k=1}^K p_k
\quad\text{s.t.}\quad p_k \ge \frac{N\gamma_{k,\min}}{\tilde m_k(\mathbf v)}\;\;\forall k.
\end{IEEEeqnarray}
One can easily shows that the closed-form power allocation solution is
\begin{equation}\label{eq:opt_power_repeat}
	p_k^*(\mathbf v)\;=\;\frac{N\gamma_{k,\min}}{\tilde m_k(\mathbf v)},\qquad k=1,\dots,K,
\end{equation}
and the total power becomes
\begin{equation}\label{eq:P_of_v}
	P(\mathbf v)\;=\;\sum_{k=1}^K \frac{N\gamma_{k,\min}}{\tilde m_k(\mathbf v)}.
\end{equation}

The placement subproblem is the minimization of $P(\mathbf v)$ over the
feasible PA positions defined earlier in \eqref{eq:op1_b} and \eqref{eq:op1_c}. Clearly, $P(\mathbf v)$ is non-convex in $\mathbf{v}$, and we use the coordinate-descent to obtain a local optimal solution.
In each
coordinate update, the $n$th PA position is optimized over its feasible interval
while the other PA positions are held fixed. The update is accepted only if it
decreases the total power. 
To mitigate sensitivity to local minima, the coordinate-descent procedure is
initialized from multiple feasible starting points. 
A summary of the proposed algorithm for the multi-antenna case is provided in Algorithm~\ref{alg:bcd_exact_repeat}.

{\bf{Remark:}} Although the proposed coordinate descent algorithm does not guarantee global optimality due to the non-convex nature of $P(v)$, it exhibits good empirical convergence behavior. In particular, the objective value is monotonically non-increasing across iterations, and the use of multiple random initializations helps mitigate the impact of poor local minima. This tradeoff between computational complexity and solution quality is common in large-scale non-convex optimization problems in wireless communications.

\begin{algorithm}[!t]
	\caption{Multi-start coordinate descent for multiple PAs}
	\label{alg:bcd_exact_repeat}
	\begin{algorithmic}[1]
		\State \textbf{Input:} $N$, $\{ \hat{\mathbf u}_k \}_{k=1}^K$, $r_k$, $L$, $\varepsilon$
		\State \textbf{Initialize} the best objective value $P_{\rm best}\leftarrow \infty$
		\For{each restart $r=1,\ldots,R$}
			\State Generate a feasible initial placement $\mathbf v^{(0)}$
			\State Set $i\leftarrow 0$
			\Repeat
				\State Compute $\hat m_k(\mathbf v^{(i)})$ for all $k$ using boundary sampling and local refinement
				\State Set $\tilde m_k(\mathbf v^{(i)})=\max\{\hat m_k(\mathbf v^{(i)}),\varepsilon\}$
				\State Set $p_k^{(i)} = N\gamma_{k,\min}/\tilde m_k(\mathbf v^{(i)})$ for all $k$
				\State Set $\mathbf v_{\rm trial}\leftarrow \mathbf v^{(i)}$ and $P_{\rm trial}\leftarrow \sum_{k=1}^K p_k^{(i)}$
				\For{$n=1$ to $N$}
					\State Determine the feasible interval of $v_n$ given the current neighbors and the waveguide constraints
					\State Solve the one-dimensional minimization of $P(\mathbf v)$ with respect to $v_n$
					\State Accept the coordinate update only if it decreases $P(\mathbf v)$
				\EndFor
				\State Recompute $P(\mathbf v_{\rm trial})$ using the updated placement
				\If{$P(\mathbf v_{\rm trial}) < P(\mathbf v^{(i)}) - \delta_{\rm tol}$}
					\State $\mathbf v^{(i+1)}\leftarrow \mathbf v_{\rm trial}$
				\Else
					\State $\mathbf v^{(i+1)}\leftarrow \mathbf v^{(i)}$
				\EndIf
				\State $i\leftarrow i+1$
			\Until{relative decrease of $P(\mathbf v)$ is below a tolerance or the maximum number of sweeps is reached}
			\If{$P(\mathbf v^{(i)}) < P_{\rm best}$}
				\State $P_{\rm best}\leftarrow P(\mathbf v^{(i)})$
				\State Store the corresponding PA placement and power allocation
			\EndIf
		\EndFor
	\State \textbf{Output:} best PA placement and corresponding power allocation
	\end{algorithmic}
\end{algorithm}

\subsection{Computational Complexity}\label{sec:complexity}

In this subsection, we analyze the computational complexity of the proposed algorithms for the single and multi-antenna cases.

For $N=1$, the robust power allocation and antenna placement problem is converted into the SDP $\mathbb{P}2$ that has $3 K + 1$ decision variables. Hence, the worst case computational complexity is $\mathcal{O}(K^3)$.

For $N > 1$, the objective function in \eqref{eq:P_of_v} depends on the worst-case channel gain $\tilde m_k(\mathbf{v})$ of each user. The computation of $\tilde m_k(\mathbf{v})$ requires $\mathcal{O}(L N)$ operations for the $L$ boundary samples and $\mathcal{O}(N_{\rm start} I_{\rm loc} N)$ operations for the local refinement stage, where $N_{\rm start}$ is the number of initial points used for refinement and $I_{\rm loc}$ denote the number of iterations required by each local worst-case refinement step (e.g., \texttt{fmincon}). Hence, the computational complexity of calculating the objective function in \eqref{eq:P_of_v} is $\mathcal{O}\left(K (L + N_{\rm start} I_{\rm loc})N\right)$. During one coordinate-descent sweep, the $N$ PA positions are updated sequentially. Each one-dimensional coordinate update invokes a line search with $I_{\rm line}$ function evaluations, so one full sweep has complexity $\mathcal{O}\left(I_{\rm line}\, K (L + N_{\rm start} I_{\rm loc}) N^2\right)$. Finally, accounting for $I_{\rm BCD}$ coordinate-descent sweeps and $N_{\rm rst}$ restarts, the overall worst-case complexity of the multi-PA algorithm is $\mathcal{O}\left(N_{\rm rst}\, I_{\rm BCD}\, I_{\rm line}\,K (L + N_{\rm start} I_{\rm loc}) N^2\right),$ which is polynomial in $N$ and linear in $K$. This represents a tradeoff between computational complexity and performance.

\section{Simulation Results}\label{sec:simulation}
Numerical results are presented in this section to evaluate the performance of the proposed schemes under both single-antenna and multi-antenna configurations. Unless otherwise specified, the simulation parameters are set as follows. The system serves $K = 3$ users, whose locations are randomly distributed within a rectangular service area of $120 \times 20~\text{m}^2$. The location uncertainty of each user is modeled as a circular region with radius $r_k = 3$~m, $\forall k$. Each user is required to achieve a target data rate of ${R}_{k, \min} = 1$~bps/Hz with a maximum allowable outage probability of $\epsilon_k = 0.01$, $\forall k$. The carrier frequency is set to $28$~GHz, and the system bandwidth is $100$~MHz. The noise power spectral density is assumed to be $-174$~dBm/Hz. The dielectric waveguide has a length of $\ell_{\mathrm{w}} = 50$~m and is deployed at a height of $d = 3$~m.

\subsection{Single-Antenna Case}

For the single-antenna case, we consider the following two benchmark schemes: 
1) a fixed-antenna baseline, where the antenna position along the $x$-axis is fixed at $-\frac{\ell_{\mathrm{w}}}{2}$, and the corresponding transmit power is determined using the geometric approach proposed in \cite{zeng2025robust} to satisfy the outage constraints; and 
2) an exhaustive-search-based upper bound, where the antenna position is optimized via exhaustive search, while the transmit power is again obtained using the geometric solution in \cite{zeng2025robust}.

\begin{figure}
\centerline{\includegraphics[width=1\linewidth]{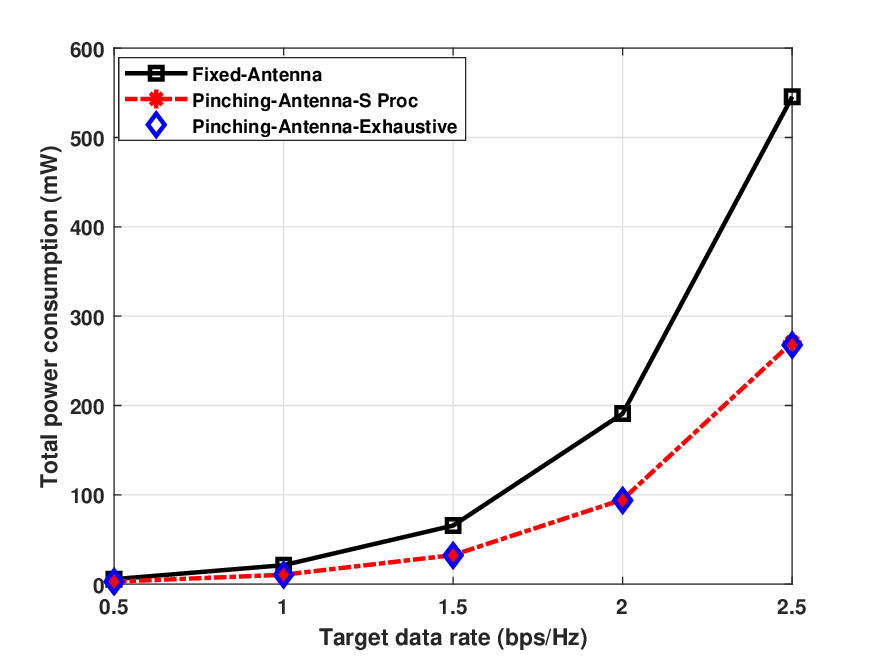}}
\caption{Total power consumption versus the target data rate at the users.} \label{rate_varies}
\end{figure}

\begin{figure}
\centerline{\includegraphics[width=1\linewidth]{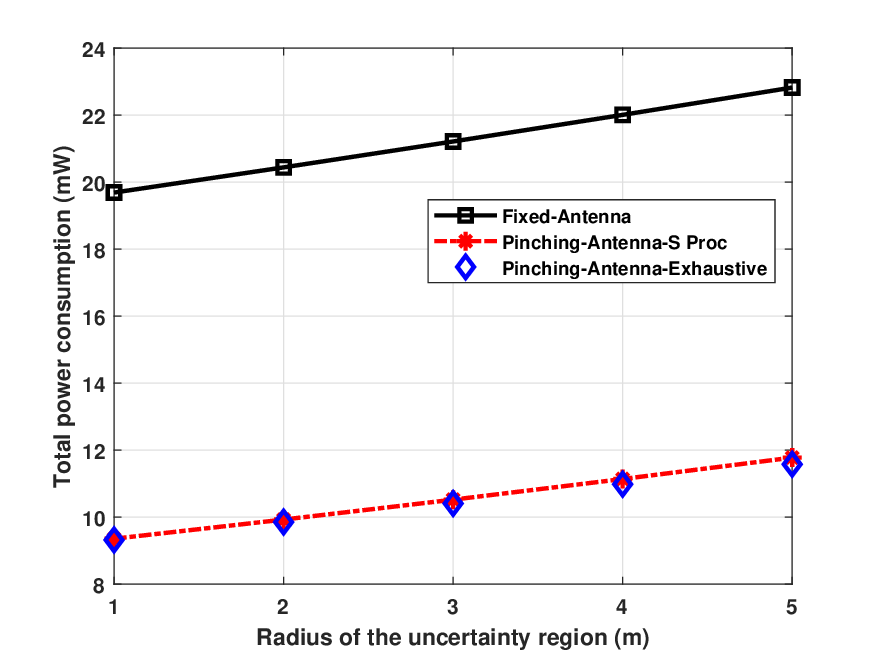}}
\caption{Total power consumption versus the uncertainty radius at the users.} \label{Radius_varies}
\end{figure}

Fig.~\ref{rate_varies} illustrates the sum transmit power as a function of the target data rate. It can be observed that, for all considered schemes, the required transmit power increases rapidly with the target rate. This behavior is consistent with the logarithmic relationship between achievable rate and SNR as dictated by Shannon’s capacity formula. Furthermore, both pinching-antenna-based schemes significantly outperform the fixed-antenna baseline, demonstrating the benefit of optimizing the antenna location. In addition, the proposed S-procedure-based robust design achieves nearly identical performance to the exhaustive-search benchmark, indicating that worst-case robustness can be attained without incurring additional power cost.

Fig.~\ref{Radius_varies} depicts the sum transmit power versus the radius of the user location uncertainty region. It is observed that the required transmit power increases approximately linearly with the uncertainty radius for all schemes. This suggests that the performance degradation due to imperfect user location information remains moderate. Notably, the proposed scheme closely matches the performance of the exhaustive-search benchmark, while achieving a substantial power reduction compared to the fixed-antenna baseline.

\begin{figure}
\centerline{\includegraphics[width=1\linewidth]{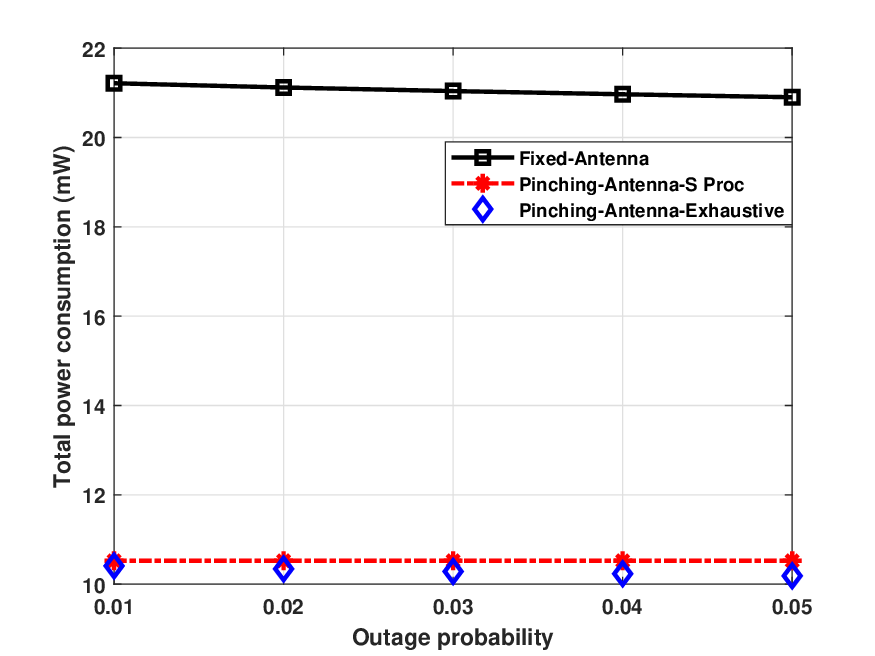}}
\caption{Total power consumption versus the outage probability constraint at the users.} \label{Outage_varies}
\end{figure}

Fig.~\ref{Outage_varies} shows the sum transmit power as a function of the allowable outage probability. Since the proposed scheme adopts a worst-case robust design that guarantees no outage within the uncertainty region, its transmit power remains invariant with respect to the outage probability, as confirmed by the simulation results. In contrast, the two benchmark schemes are based on probabilistic constraints, and thus their required transmit power decreases as the outage tolerance increases. However, this reduction is relatively marginal, particularly for the exhaustive-search benchmark. Even when $\epsilon_k = 0.05$, the performance gap between the proposed scheme and the exhaustive-search solution remains negligible, demonstrating that worst-case robustness can be achieved with minimal power increase.

\begin{figure}
\centerline{\includegraphics[width=1\linewidth]{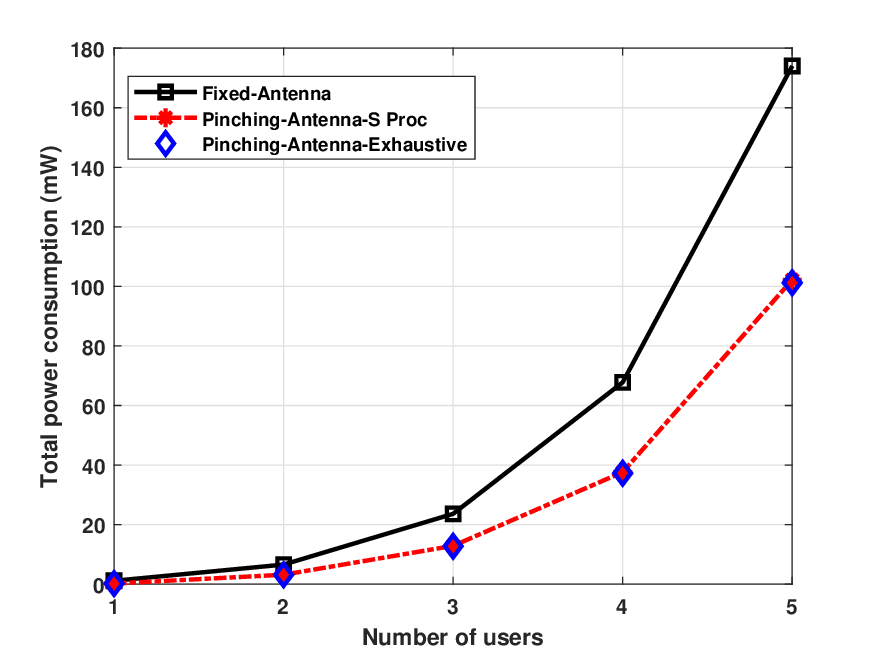}}
\caption{Total power consumption versus the number of users; ${R}_{k, \min} = 1$~bps/Hz.} \label{Number_varies}
\end{figure}

Fig.~\ref{Number_varies} illustrates the sum transmit power as a function of the number of users. It can be observed that the required transmit power increases rapidly with the number of users for all considered schemes, exhibiting an exponential growth trend similar to that observed when increasing the minimum rate requirement. This behavior can be explained by the expression $\gamma_{k,\min} \triangleq \sigma_k^2 \big(2^{K R_{k,\min}} - 1\big)$, which shows that increasing the number of users $K$ has a similar impact on $\gamma_{k,\min}$ as increasing the target rate $R_{k,\min}$. 
Furthermore, the fixed-antenna scheme requires significantly higher transmit power compared to the two pinching-antenna-based schemes, which exhibit nearly identical performance and overlap across the considered range.

\subsection{Multi-antenna Case}
We now extend the analysis to the multi-antenna scenario. The proposed pinching-antenna-based solution described in Section III-B is referred to as ``Pinching-Antenna-CD.'' As a benchmark, we consider a fixed-antenna scheme in which $N$ antennas are uniformly deployed along the waveguide around $-\frac{\ell_{\mathrm{w}}}{2}$ with an inter-element spacing of $\frac{\lambda}{2}$. For this scheme, the total transmit power is computed according to \eqref{eq:worstcase_def_repeat}--\eqref{eq:P_of_v}.

\begin{figure}
\centerline{\includegraphics[width=1\linewidth]{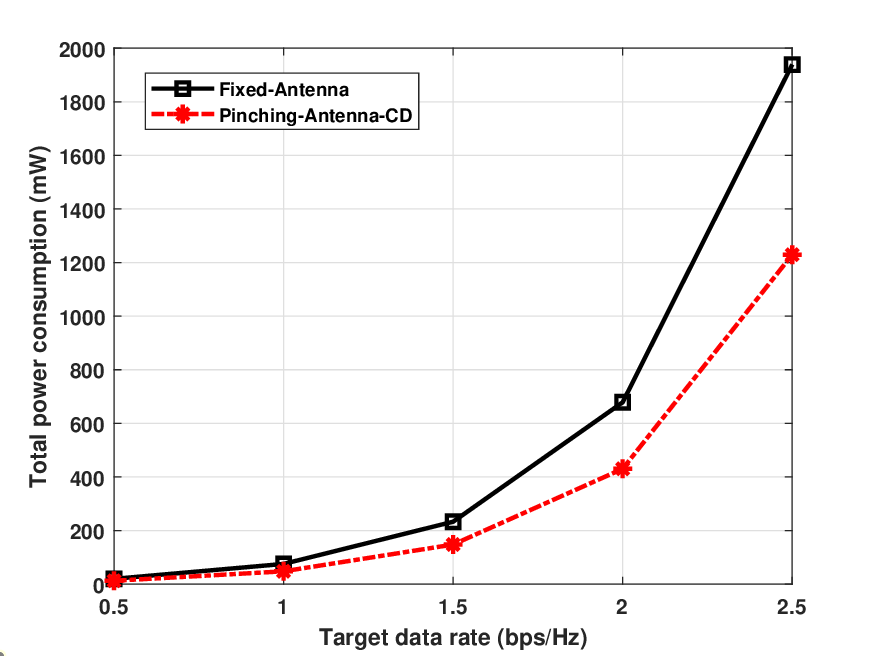}}
\caption{Total power consumption versus the target data rate at the users; $N=3$, $K=3$ and $r_k=3$ m.} \label{rate_varies_2}
\end{figure}

\begin{figure}
\centerline{\includegraphics[width=1\linewidth]{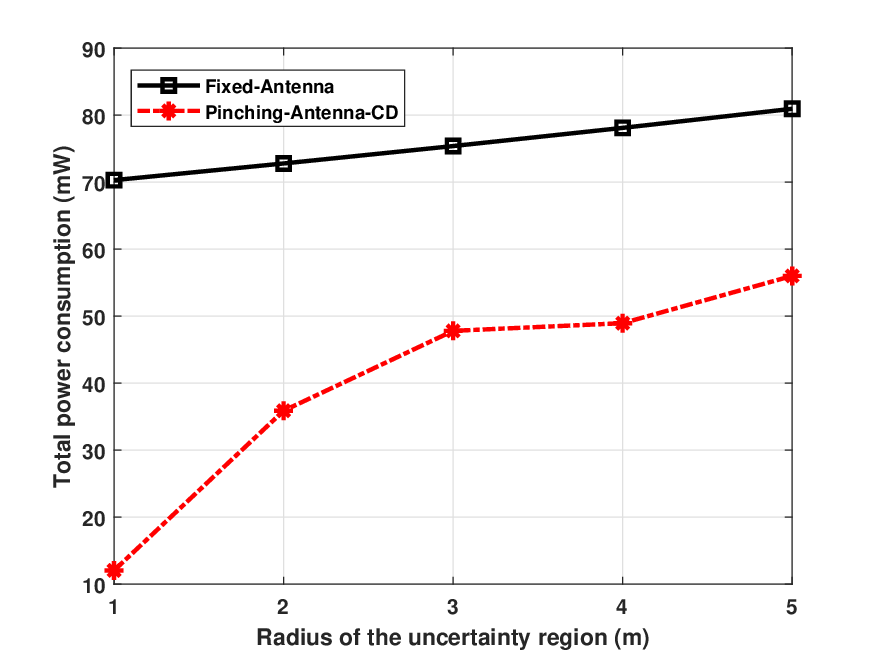}}
\caption{Total power consumption versus the uncertainty radius at the users; $N=3$, $K=3$ and ${R}_{k, \min} = 1$~bps/Hz.} \label{Radius_varies_2}
\end{figure}

Fig.~\ref{rate_varies_2} depicts the total transmit power as a function of the target data rate for $N=3$ antennas. Similar to the single-antenna case, the required transmit power increases rapidly with the minimum rate requirement, which is consistent with the exponential dependence of the required SNR on the target rate. Furthermore, the proposed pinching-antenna scheme consistently outperforms the fixed-antenna benchmark across all considered rate values, highlighting the advantage of jointly optimizing antenna positions in multi-antenna systems. 

Fig.~\ref{Radius_varies_2} shows the total transmit power versus the radius of the user location uncertainty region. For the fixed-antenna scheme, the required power increases approximately linearly with the uncertainty radius, which aligns with the observations in the single-antenna case. In contrast, the proposed scheme exhibits a non-monotonic but overall increasing trend. This behavior is attributed to the coupling between antenna position optimization and worst-case channel conditions under uncertainty. Nevertheless, the proposed scheme achieves a substantial power reduction compared to the fixed-antenna baseline across all uncertainty levels.

From a system design perspective, these results highlight that antenna position optimization becomes increasingly important in the presence of uncertainty. In particular, while fixed antenna deployments are unable to adapt to unfavorable channel realizations, PA systems can dynamically reposition radiating elements to mitigate worst-case conditions.

\begin{figure}
\centerline{\includegraphics[width=1\linewidth]{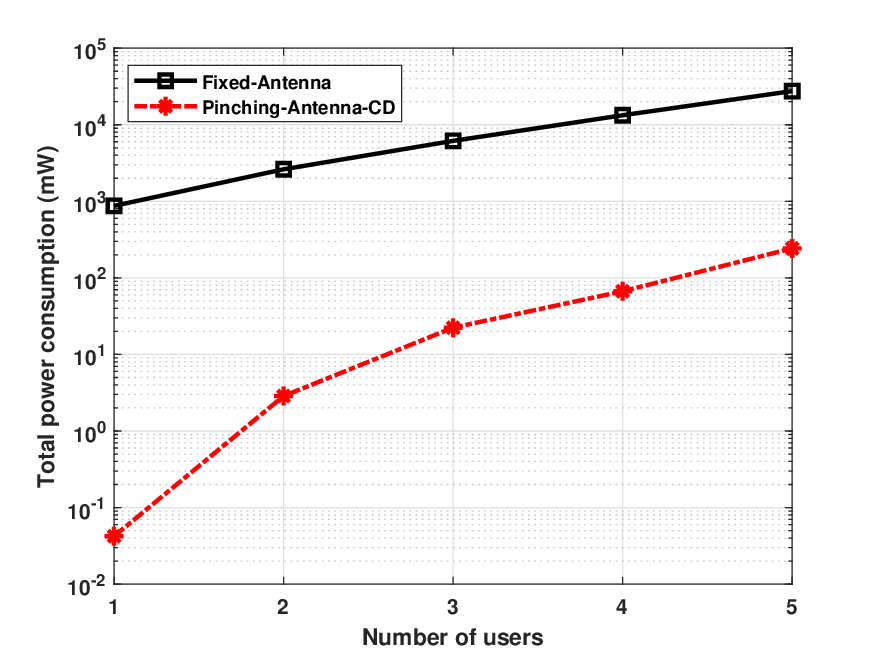}}
\caption{Total power consumption versus the number of users; $N=3$, $r_k=3$ m and ${R}_{k, \min} = 1$~bps/Hz..} \label{Number_varies_2}
\end{figure}

Fig.~\ref{Number_varies_2} illustrates the total transmit power as a function of the number of users. Due to the significant performance gap between the two schemes, the $y$-axis is presented in logarithmic scale. For the fixed-antenna benchmark, the total transmit power increases sharply with the number of users, reflecting an exponential growth trend similar to that observed in the single-antenna case. The proposed pinching-antenna scheme follows a similar trend but with significantly lower power consumption. For instance, at $K=5$, the fixed-antenna scheme requires more than $2.7\times 10^4$ mW, whereas the proposed scheme only requires $243$ mW, demonstrating a substantial performance gain.

\begin{figure}
\centerline{\includegraphics[width=1\linewidth]{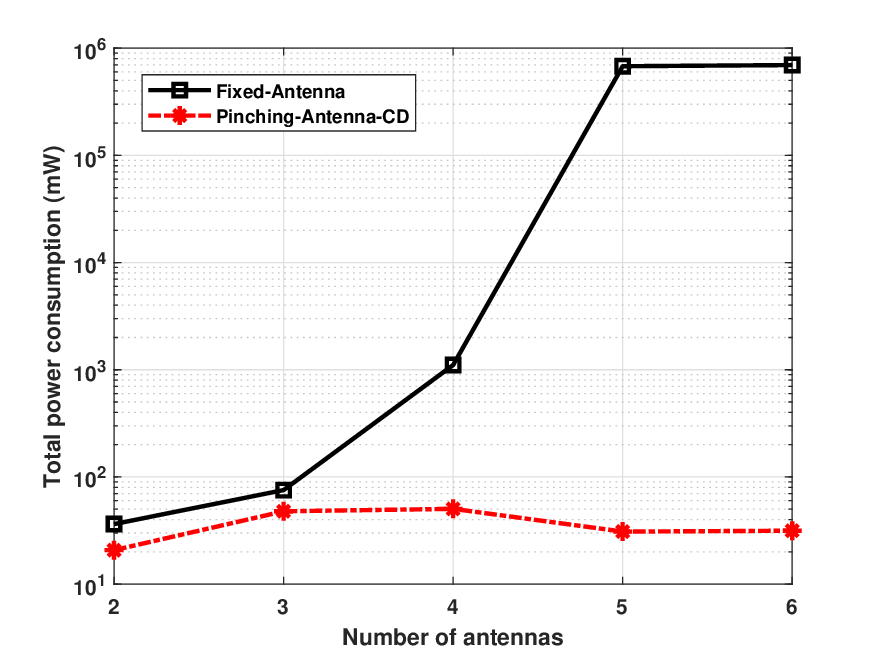}}
\caption{Total power consumption versus the number of pinching antennas; $K=3$, $r_k=3$ m and ${R}_{k, \min} = 1$~bps/Hz.} \label{antenna_varies}
\end{figure}

Finally, Fig.~\ref{antenna_varies} presents the total transmit power as a function of the number of antennas. A logarithmic scale is again adopted for clarity. For the fixed-antenna scheme, the total transmit power increases rapidly with the number of antennas, reaching approximately $6.7 \times 10^5$ mW at $N=5$. This counterintuitive behavior stems from the worst-case robust design: with fixed antenna positions, increasing the number of antennas intensifies signal superposition effects, which can exacerbate unfavorable channel realizations within the uncertainty region and thus require higher transmit power to guarantee the rate constraints.

In contrast, the proposed pinching-antenna scheme effectively mitigates this issue by optimizing antenna positions. As a result, the transmit power does not monotonically increase with $N$; instead, it first increases and then decreases as additional spatial degrees of freedom are exploited to better adapt to user locations and uncertainty. The performance gap between the two schemes becomes increasingly pronounced for larger $N$, clearly demonstrating the benefit of antenna position optimization in multi-antenna pinching systems.

The non-monotonic behavior observed for the proposed scheme suggests that simply increasing the number of antennas is not always beneficial unless accompanied by proper spatial optimization. This observation provides an important design guideline: in PA systems, the spatial configuration of antennas is as critical as their number.

\subsection{Discussion}
The proposed robust design framework provides several important insights for practical PA system deployment. First, worst-case robust optimization offers a reliable alternative to probabilistic approaches when uncertainty distributions are unknown or difficult to estimate. Second, the results demonstrate that antenna position optimization plays a more critical role than power allocation alone, particularly in multi-antenna settings. 

Moreover, the proposed framework can be extended to incorporate additional system constraints, such as hardware impairments and multi-user interference. These extensions represent promising directions for future research and could further highlight the flexibility of PA systems as a reconfigurable wireless architecture.

\section{Conclusion}
This paper investigated robust power allocation and antenna placement for pinching antenna systems under user location uncertainty. By modeling the uncertainty as bounded regions, a worst-case robust framework was developed for both single-antenna and multi-antenna configurations. For the single-antenna case, the joint optimization problem was reformulated as a convex SDP via the S-procedure, enabling efficient computation of the globally optimal solution without relying on heuristic search methods. For the multi-antenna case, the inherent non-convexity introduced by channel superposition was addressed through an efficient worst-case channel gain evaluation method combined with a block coordinate descent algorithm for antenna placement.
Simulation results demonstrated that the proposed robust designs provide strong reliability against location uncertainty while achieving transmit power performance close to that of outage-based benchmark schemes. In particular, the results highlight that worst-case robustness can be attained with negligible additional power cost, and that optimizing antenna positions is critical for unlocking the full potential of multi-antenna PA systems. 

\bibliographystyle{IEEEtran}
\bibliography{biblio}

\end{document}